\documentclass[prd,superscriptaddress,amsmath,notitlepage,twocolumn]{revtex4-1}
\usepackage{graphicx}
\usepackage{float}
\usepackage{color}
\usepackage{epstopdf,cancel,ulem}
\usepackage{epsf,latexsym,bbm,euscript}
\usepackage{amssymb,amsmath}
\usepackage{mathtools} 
\usepackage{times,graphics}
\usepackage{soul,xcolor}
\usepackage{epsfig}
\usepackage[T1]{fontenc}
\usepackage{booktabs,appendix}
\usepackage{multirow,bigdelim}
\usepackage{cancel}
\usepackage{url,hyperref}
\hypersetup{colorlinks,linkcolor={blue!55!black},citecolor={red!50!black},urlcolor={blue!45!black}}

\def\6{{\langle}}
\def\9{{\rangle}}

\newcommand{\be}{\begin{equation}}
\newcommand{\ee}{\end{equation}}
\newcommand{\ba}{\begin{eqnarray}}
\newcommand{\ea}{\end{eqnarray}}

\newcommand{\beq}{\begin{equation}}
\newcommand{\eeq}{\end{equation}}
\newcommand{\beqa}{\begin{eqnarray}}
\newcommand{\eeqa}{\end{eqnarray}}

\def\be{\begin{equation}}
\def\ee{\end{equation}}

\def\bali{\begin{align}}
\def\eni{{\end{align}}}

\def\1{{{\mathbbm 1}}}

\def\half{{\tfrac{1}{2}}}

\def\pad{{\partial}}

\def\sg{\textsl{g}}

\def\cO{\mathcal{O}}

\begin{document}

\title{Energy-momentum tensor and metric near the  Schwarzschild sphere}

\author{Valentina Baccetti}
\affiliation{Department of Physics and Astronomy, Macquarie University, Sydney NSW 2109, Australia}
\affiliation{School of Science, RMIT University, Melbourne, VIC 3000, Australia}
\author{Robert B. Mann}
\affiliation{Department of Physics and Astronomy, University of Waterloo, Waterloo, Ontario N2L 3G1, Canada}
\affiliation{Perimeter Institute for Theoretical Physics, Waterloo, Ontario N2L 6B9, Canada}
\author{Sebastian Murk}
\author{Daniel R. Terno}
\affiliation{Department of Physics and Astronomy, Macquarie University, Sydney NSW 2109, Australia}

\begin{abstract}
	Regularity of {the horizon radius $r_\sg$ of a collapsing body} constrains a limiting form of a spherically symmetric energy-momentum tensor near it. Its nonzero limit belongs to one of four classes that are distinguished only by two signs. As a result, close to $r_\sg$  the geometry can always be described by either an ingoing or outgoing Vaidya metric with increasing or decreasing mass. If according to a distant outside observer the trapped regions form in finite time, then the Einstein equations imply violation of the null energy condition. In this case the horizon radius and its rate of change determine the metric in its vicinity, and the hypersurface $r=r_\sg(t)$ is timelike during both the expansion and contraction of the trapped region. We present the implications of these results for the firewall paradox and discuss arguments that the required violation of the null energy condition is incompatible with the standard analysis of black hole evaporation.
\end{abstract}
\maketitle

\section{Introduction}

Astrophysical black holes (ABH) \cite{fn:book,abh,cp:na17} ---  massive compact dark objects ---  are commonly found in the observable universe. 
Current data are consistent with having the Schwarzschild or Kerr-Newman solutions of classical general relativity (GR) as asymptotic final states of the collapse \cite{abh,cp:na17,visser}. Nevertheless, the question of if, how and when ABHs develop any of the horizons and/or singularities that are predicted by GR \cite{fn:book,abh,cp:na17,visser,he:book} is still open \cite{abh,cp:na17}.

  Quantum effects \cite{fn:book,bd:82,bmps:95,visser,rbm:book,qua-mod,hay:06,fg:11,coy:15} add additional complexity. Event horizons may become optional \cite{visser,hay:06,blsv}, and the notion of a black hole (BH) is then tied with one of the locally or quasilocally defined surfaces, such as an apparent horizon \cite{faraoni:13,krishnam:14}. Hawking radiation accompanies the formation and evolution of black holes \cite{haw:74}. Its explicit form was originally obtained on the background of eternal black holes that are the vacuum solutions of GR \cite{fn:book,bmps:95}. Later it was shown  {that} this phenomenon does not require formation of an event or even  of an  apparent horizon \cite{visser,blsv,pp:09}. If the radiation is not terminated when the collapsing object reaches some macroscopic scale or becomes a Planck-scale remnant \cite{coy:15}, the picture of an ABH as being in a permanent state of asymptotic approach to horizon formation becomes untenable as the object itself disappears in a finite time. Hawking radiation violates all energy conditions and thus provides additional possibilities both for having singularity-free objects with or without trapped regions, as well as naked singularities \cite{cp:na17,coy:15}.

Given that there are several related horizon  definitions  that  are based on bounding the region of negative expansion of the outgoing null
 geodesics emanating from a spacelike compact two-dimensional surface with spherical topology, the definition of a physical BH (and thus the question of their very existence)
 is somewhat fuzzy. 
Hence we use the term ``black hole" to designate any massive dense object that contains a trapped spacetime region, regardless of the presence of a suitably defined event horizon. They are observationally relevant physical objects (and not just useful mathematical idealizations) if formed at the finite time of a distant observer.

Our starting point is that such a BH exists; i.e., a trapped region is formed in a finite time $t_\mathrm{S}$ of Bob. We do not assume  that this process is accompanied by Hawking radiation. If quantum effects lead to BH evaporation, our only assumption is that the evaporation time $t_\mathrm{E}>t_\mathrm{S}$. Our goal is to explore the implications of a compact object actually being a BH.


 We assume validity of semiclassical gravity \cite{pp:09,bmt-1} 
 and describe dynamics via the Einstein equations where the standard curvature terms 
are equated to the expectation value of the renormalized stress-energy tensor. It represents the entire matter content of the model: both the collapsing matter and the created
excitations of the quantum fields are included.
This cumulative representation allows a self-consistent study of the dynamics without having recourse to the usual iterative calculations of the backreaction \cite{qua-mod}.

 We shall restrict our considerations to spherical symmetry,  set $\hbar=c=G=1$ and use the $-+++$ signature.

 \section{Tensors near the Schwarzschild sphere}\label{se2}

Here we demonstrate how regularity of the apparent horizon constrains the metric in its vicinity. Moreover, if it is formed in a finite time of a distant observer, the family of allowed metrics is essentially unique.

\subsection{General considerations}

 The most general spherically symmetric metric in $(z,r)$ coordinates, where   $z$ is either
  the Schwarzschild time $t$ or the advanced or   retarded null coordinate  $u_\pm$,  is

 \begin{align}
 ds^2&=-e^{2h(t,r)}f(t,r)dt^2+f(t,r)^{-1}dr^2+r^2d\Omega \label{sgenm}\\
 &=-e^{2h_\pm(u_\pm,r)}f_\pm(u_\pm,r)du^2_\pm\pm 2e^{h_\pm(u_\pm,r)}du_\pm dr+r^2d\Omega.
 \end{align}
 The function $f$ is coordinate independent \cite{bardeen:81,blau}, i.e., $f(t,r)=f_\pm(u_\pm(t,r),r)$, and we can decompose it as
\be
f=1-C(t,r)/r=1-C_\pm(u_\pm,r)/r.
\ee
 The functions $h$ and $h_\pm$ play the role of integrating factors \cite{blau} that turn, e.g., the expression
\be
dt=e^{-h}(e^{h_\pm}du_\pm\mp f^{-1}dr), \label{intf}
\ee
 into an exact differential, provided that the coordinate transformation exists.

 In an asymptotically flat spacetime $h\to 0$ and $f\to 1$ as $r\to\infty$, and $t$ is the physical time of a stationary Bob at   spacelike infinity.
 In the Schwarzschild spacetime $C=2M$, and the coordinates $u_\pm$  are the retarded $u_-\equiv u \vcentcolon= t-r_*$ and advanced $u_+\equiv v \vcentcolon= t+r_*$  Eddington-Finkelstein coordinates,
where $r_*$ is the tortoise coordinate.

On the other hand, having
$h\to 0$ and $f\to 1- r^2/R_H$, where $R_H=1/H_\Lambda$ is the Hubble radius, asymptotically describes  a flat de Sitter space or a closed de Sitter space in static coordinates \cite{muk}.
 An explicit example of a
gravitationally bound system in an expanding homogeneous and isotropic universe is furnished by the McVittie metrics \cite{mv-m},  where late-time exponential
expansion results    in a de Sitter-Schwarzschild BH.

Regions of nonpositive expansion of outward pointing future-directed radial null geodesics  (in asymptotically flat spacetimes) exist only if the equation $f(z,r)=0$
has a root \cite{krishnam:14,blau}.
This root (or, if there are several, the largest one) is the  {Schwarzschild horizon radius} $r_\sg(z)$.  Our first assumption is the existence of  $r_\sg(t)$  for some period of time
beginning at $t_\mathrm{S}<\infty$.
 If the spacetime is not asymptotically flat, we assume the separation of scales.  {In the de Sitter space example above, having stationary Bob's radial coordinate
 satisfy $r_\sg\ll r \ll R_H$, i.e., positioning him
 far outside the  {horizon radius}    and well inside the Hubble radius, puts him into an effectively flat region where $t$ is the physical time.}

 Regularity of the hypersurface $r=r_\sg$ is the standard assumption in the {semiclassical} theory \cite{abh,fn:book,rbm:book}.  Ricci and  Kretschmann  scalars have been used
 in   investigations of BH evaporation without an event horizon \cite{bh-nohor}. We choose two scalars that are expressed directly in terms of the energy-momentum tensor
 $T_{\mu\nu}$,
  namely, its trace
 $T \vcentcolon= T^\mu_{~\mu}\equiv R/8\pi$, where $R$ is the Ricci curvature scalar, and the square $\mathfrak{T} \vcentcolon= T^{\mu\nu}T_{\mu\nu}$.
Our second assumption is that these scalars may have  arbitrary but finite values,
   $|T(z,r_\sg)|<\infty$ and $|\mathfrak{T}(z,r_\sg)|<\infty$.

   Finite curvature  does not imply that the functions $h$ and $C$ in a particular coordinate system are finite. However, from the definition of $r_\sg$ it follows that
\be
 C (t,r)=r_\sg(t)+W(t,x),  \qquad W(t,0)=0,
 \ee
where $x \vcentcolon= r-r_\sg(t)$ and $W(t,x)<x$ for $x>0$. Derivatives of $W$ can diverge on the approach  to $x=0$, but only moderately so as to ensure its continuity.

The  Einstein equations that determine the functions $h$ and $C$  are
\begin{align}
G_{tt}=&\frac{e^{2h}(r-C)\pad_r C}{r^3}=8\pi T_{tt}, \label{gtt}\\
G_t^{\,r}=&\frac{\pad_t C}{r^2}=8\pi T_t^{\,r}, \label{gtr}\\
G^{rr}=&\frac{(r-C)(-\pad_r C+2(r-C)\pad_r h)}{r^3}=8\pi T^{rr}. \label{grr}
\end{align}
This is the simplest form of the equations. It provides a natural choice of the independent components of the energy-momentum tensor.
The metric of Eq.~\eqref{sgenm} entails  $T^\theta_{~\theta}\equiv T^\phi_{~\phi}$. {Then} the trace and the square scalars of the energy-momentum tensor are
 \begin{align}
    T=&-e^{-2h }T_{tt}/f  +T^{rr}/f +2T^\theta_{~\theta}, \label{fin1}\\
  \mathfrak{T}=&-2\left(\frac{e^{-h} T_t^{\,r}}{f }\right)^2+\left(\frac{e^{-2h }T_{tt}}{f }\right)^2  +\left(\frac{T^{rr}}{f }\right)^2 +2\big(T^\theta_{~\theta}\big)^2.       \label{fin2}
    \end{align}
For   future convenience we introduce $\tau_t \vcentcolon= e^{-2h }T_{tt}$, $\tau^r \vcentcolon= T^{rr}$ and $\tau_t^r \vcentcolon= e^{-h} T_t^{\,r}$.

Regularity of {the invariants $T$ and $\mathfrak{T}$}
holds independently of the left-hand side of the Einstein equations, which may involve quantum corrections \cite{bd:82}.
 Without any additional information  we have to assume that the  component
$T^\theta_{\,\theta}$ is finite at $r_\sg$.  If the dynamics is described by the Einstein equations, this intuitive property follows from the consistency of Eqs.~\eqref{gtt}--\eqref{grr} (see Appendix \ref{apreg}).

The regularity of
$T$ and $\mathfrak{T}$
 at the Schwarzschild sphere implies that either the three limits of  $\tau_t$, $\tau^r$ and $\tau_t^r$ are jointly zero as $r\to r_\sg$,  approaching it at least
 as fast as $f(t,r)$, or  the divergences in Eqs.~\eqref{fin1} and \eqref{fin2} cancel out, so
\be
\lim_{r\to r_\sg} \tau^r=\lim_{r\to r_\sg} \tau_t=\Xi(t), \qquad \lim_{r\to r_\sg} \tau_t^r=\pm \Xi(t),  \label{limits} \ee
for some function $\Xi(t)$. {Indeed, from Eq.~\eqref{fin1} it follows that
\be
\lim_{r\to r_\sg} T=\lim_{r\to r_\sg}( -\tau_t+\tau^r)/f +2T^\theta_{~\theta}
 \ee
is finite, and thus $\tau_t\to \tau^r$ faster than $f\to 0$, and Eq.~\eqref{fin2} implies  $|\tau_t^r|\to|\tau_r|$. The form of the function $\Xi(t)$ is  not constrained by regularity considerations.}

We consider in detail the case $\Xi(t)\neq 0$, as it is more general and {existing explicit calculations of $\6\hat T_{\mu\nu}\9$ indicate that it is present
\cite{fn:book,bmps:95,leviori:16}.  }  The case of all three limits of $\tau_t$, $\tau^r$ and $\tau_r^t$ being $\Xi =\vcentcolon -\Upsilon^2$   is supported by the explicit calculations  
of the renormalized energy-momentum tensor for Unruh vacuum on the background of an eternal
black hole (in that case $h\equiv 0$)   \cite{leviori:16}, where  the components $T^{rr}$, $T_{tt}$ and $T^t_r$ were shown to approach the same limit as $r\to r_\sg$. 
 {Appendix \ref{apXi}} provides the details, as well as the analysis of the case $\Xi=0$ that leads to the same qualitative conclusions.  We also consider here only the case
 of $T_{\mu\nu}$ that does not contain additional (milder) singular terms, as their presence does not change the leading terms in the solutions below.
  Close to $r_\sg$
 Eq.~\eqref{gtt}  becomes
 \be
   \pad_x W \approx \frac{8\pi\Xi r_\sg^3}{x-W}. \label{Wpos}
   \ee
Due to the singularity at $x=0$  Eq.~\eqref{Wpos} with the initial condition $W(t,0)=0$ has two real solutions in terms of   Lambert functions \cite{lambert}. Both require
 $\Xi=-\Upsilon^2<0$ for some $\Upsilon(t)>0$.    We will see below that this implies violation of
the null energy condition (NEC) \cite{he:book, mmv:17}. This is consistent with the well-known result  that a trapped surface cannot be ``visible'' from   future null
infinity $\mathcal{I}^+$ unless a
weak energy condition is violated \cite{fn:book,he:book}. However, here it is
a  local result that is valid even if   $\mathcal{I}^+$ is not defined.

The auxiliary condition  $W(t,x)<x$ selects the solution
in terms of the Lambert function $\mathsf{W}_{-1}$. Since the equation is approximate, we should use only the leading part  of the solution
 that can also be obtained by direct evaluation of the series.   It is
 \be
W= -4\Upsilon\sqrt{\pi r_\sg^3 x}+\frac{1}{3}x\ldots,
\ee
and it coincides with the leading part of the solution of the exact
Einstein equation \eqref{gtt}. In terms of  $\alpha^2 \vcentcolon= 16 \pi \Upsilon^2 r_\sg^3$, the mass function becomes
 \be
C=\frac{2}{3}r_\sg(t)-\alpha(t)\sqrt{r-r_\sg(t)}+\frac{1}{3}r\ldots.  \label{c0sin}
\ee
 Substitution   into Eq.~\eqref{grr} leads to
\be
\partial_x h \approx -\frac{8\pi \Upsilon^2 r_\sg^3}{(x-W)^2} \approx - \frac{\alpha^2}{2\left(\alpha\sqrt{x} +2x/3\right)^2}.
\ee
Its  exact solution is
\be
h=h_0(t)-\ln\frac{\sqrt{x}}{3\alpha+2 \sqrt{x}}-\frac{3\alpha}{3\alpha+2\sqrt{x}} =\vcentcolon h_0(t)+h_1(t,x).
\label{h1}
\ee
 The function of time $h_0(t)$ is  not determined by the Einstein equations. Its choice determines the choice of the time variable.
Higher-order
terms depend on the exact form of $\tau_t$ and $\tau^r$.  Hence
\be
h=-\ln\frac{\sqrt{x}}{\xi_0(t)} +\frac{4}{3\alpha}\sqrt{x}+\cO(x),   \label{h2}
\ee
where the function $\xi_0(t)$ is determined by the choice of the time variable.  For example, expansion of Eq.~\eqref{h1} leads to $\xi_0=3\alpha e^{ {h_0-1}}$.

The flux sign is determined {by} the sign of $\pad_t C(t,r)$.
Since \eqref{gtr} has to be consistent with \eqref{c0sin},
we match the singular part of $\pad_tC$ that is obtained from  Eq.~\eqref{gtr},
\be
\frac{\pad_t C  }{r^2}=8\pi T_t^{\,r} {\approx} \pm 8\pi\Upsilon^2e^{h} = \pm\frac{8\pi \Upsilon^2 {\xi_0}}{\sqrt{r-r_\sg}}+\ldots,      \label{ert}
\ee
with the singular part of
\be
 \pad_t C =\dot r_\sg+\pad_t W=\frac{2 \Upsilon\sqrt{\pi r_\sg^3}}{\sqrt{r-r_\sg}} \dot r_\sg+\ldots,
\ee
 where the limit of the omitted terms is zero. As a result
 \be
 \dot r_\sg/{\xi_0}=\pm4\sqrt{\pi}\,\Upsilon\sqrt{ r_\sg}= \pm \alpha/r_\sg.       \label{lumin}
 \ee
A direct calculation  confirms that  the scalars $R=R^\mu_{\,\mu}$ and $\mathfrak{R}=R^{\mu\nu}R_{\mu\nu}$   are indeed finite, as is the
Kretschmann  scalar $K=R^{\mu\nu\lambda\rho}R_{\mu\nu\lambda\rho}$.  For example, the  Ricci scalar can be expanded as
\be
R=\varrho_3 x^{-3/2}+\varrho_2 x^{-1}+\varrho_1 x^{-1/2}+\cO(x),
\ee
where the coefficients of the divergent terms are
\be
\varrho_3=(\dot r_\sg^2-16\pi \xi_0^2r_\sg\Upsilon^2)\rho_3,
\ee
and
\begin{align}
\varrho_2& =(\dot r_\sg^2-16\pi \xi_0^2r_\sg\Upsilon^2)\rho_2, \\
\varrho_1&=\rho_1\big(-32\pi \xi_0^3 r_\sg\Upsilon^2-18\pi\dot\xi_0\dot r_\sg r_\sg^3\Upsilon^2 \nonumber \\
&+\xi_0[(2-9\pi r_\sg^2\Upsilon^2)\dot r_\sg^2-18\pi r_\sg^3\Upsilon(\dot r_\sg\dot \Upsilon-  {\Upsilon}\ddot r_\sg)]\big),
\end{align}
with the explicit form of the functions $\rho_{1,2,3}$ given in Appendix\,\ref{apR}, Eq.~\eqref{rhoR}.

The coefficients of $\varrho_3$ and $\varrho_2$ are identically zero due to Eq.~\eqref{lumin}, and the coefficient $\varrho_1$ is
 identically zero due to Eq.~\eqref{lumin} and its derivative that relates $\dot\Upsilon$
and $\ddot r$.

 At a fixed time   qualitative changes in both the $h$ and $C$ functions occur at $x\sim \alpha^2 =\dot r_\sg^2 r_\sg^2/\xi_0^2$. For the distances $\alpha^2\lesssim x\ll r_\sg$ we have
 \be
 C(t,x)\approx r_\sg(t)+x/3.
 \ee

\subsection{Asymptotically flat spacetime}
 A more explicit expression for $\xi_0$ can be obtained in an asymptotically flat spacetime. In this case the time $t$ is the physical time at spacelike infinity, hence $\lim\limits_{x\to\infty} h(t,x)=0$, while
 \be
 h(t,x)=g_0(t)-\half \ln x +\cO(\sqrt{x}),
 \ee
for some function $g_0(t)$ for $x \to 0$ is satisfied by all spherically symmetric solutions with a finite-time formation of an apparent horizon.

We can decompose
\be
h(t,x)=h_1(t,x)+h_2(t,x),
\ee
where
\be
\lim_{x\to \infty} h_2(t,x)=-\lim_{x\to \infty}h_1(t,x)=-\ln 2.
\ee

Series expansion as in Eq.~\eqref{h2} indicates that the higher-order terms in the energy-momentum tensor contribute only to the terms of the order $x$ or higher. On the other hand, this expansion does not set
the limit of $h_2$ when $x\to 0$. If we can assume that it is zero, then
\be
g_0=\ln (3\alpha)-1. \label{g0sim}
\ee
In a general asymptotically flat spacetime  Eq.~\eqref{gtr} becomes
\be
\frac{\pad_t C  }{r^2}= \pm\frac{8\pi \Upsilon^2e^{g_0}} {\sqrt{r-r_\sg}}+\ldots,      \label{ert}
\ee
and if Eq.~\eqref{g0sim} applies,
\be
\dot r_\sg=\pm\frac{3\alpha^2}{r_\sg \,e}.
\ee

\subsection{Retarded and advanced coordinates}

Null coordinates $u_\pm$  allow for simpler metric expressions. In the case $\tau_t^r=-\Upsilon^2$  Eq.~\eqref{intf} leads to a particularly simple set of the Einstein equations in $(u_+,r)$ coordinates
in the vicinity of $r_\sg=r_+$:
\begin{align}
&\pad_{u_+} C_+=-8\pi r_+^2e^{h_+}\Upsilon^2+\cO(r-r_+), \\ \label{negvev}
&\pad_r C_+=\cO(r-r_+), \qquad  \pad_r h_+=\cO(r-r_+). \end{align}
 Their solution is regular,
 \be
 C_+=r_+(u_+)+\cO(r-r_+)^2, \qquad h_+=\cO(r-r_+)^2,
 \ee
where $dr_+/du_+=-8\pi r_+^2\Upsilon^2$, and the coefficients of the higher-order terms are determined by the full energy-momentum tensor. Then close to the apparent horizon
\be
ds^2=-\big(1-r_+/r\big)du_+^2+2du_+dr+r^2d\Omega. \label{metdec}
\ee
This form of the metric --- ingoing Vaidya with decreasing mass --- agrees with the near-horizon form of the metric of an evaporating BH \cite{bmps:95}.

 Similarly,  in the case $\tau_t^r=+\Upsilon^2$  the coordinates $(u_-,r)$ are particularly convenient. The Einstein equations become
\begin{align}
&-\pad_{u_-} C_-=-8\pi r_+^2e^{h_-}\Upsilon^2+\cO(r-r_-), \\ \label{mequeu}
&\pad_r C_-=\cO(r-r_-), \qquad  \pad_r h_-=\cO(r-r_-),
\end{align}   resulting in  the limiting form of the metric
\be
ds^2=-\big(1-r_-/r\big)du_-^2-2du_-dr+r^2d\Omega, \label{metin}
\ee
which is the outgoing Vaidya metric with increasing mass.

          \section{Null energy condition near the Schwarzschild sphere}

The  limiting form of the $(tr)$ block of $T_{\mu\nu}$ is \be
T^a_{~b}=\begin{pmatrix}
-\Xi/f & -s e^{-h}\Xi/f^2 \vspace{1mm}\\
s e^h  \Xi & \Xi/f
\end{pmatrix},     \quad
 T_{\hat{a}\hat{b}}=  \frac{\Xi}{f}   \begin{pmatrix}
1 & s  \vspace{1mm}\\
s   & 1 \end{pmatrix},
  \label {tneg}
\ee
where $s=\pm 1$ and the second expression is written in the orthonormal frame.
This result is independent of the Einstein equations and remains valid even if higher-order curvature terms are added to their left-hand side.

For $\Xi=-\Upsilon^2$ and $s=\pm 1$  the NEC is violated;  $T_{\hat{a}\hat{b}}k^{\hat a}k^{\hat b }<0$
for {a} radial null vector $k^{\hat a}=(1,s,0,0)$. This is the case we have considered in Sec.~\ref{se2}.
 Using the metric in $(u_\pm,r)$ coordinates we see that the Schwarzschild sphere is the outer boundary of   spherically symmetric marginally trapped surfaces.

  Unlike  models of collapse that
satisfy energy conditions,
the finite-time Schwarzschild sphere is timelike  {during both} its growth and shrinking. Indeed,  in the growth phase we parametrize it as a hypersurface
\be
\Phi(x)=r-r_-(u_-)=0.
\ee
{Then the  normal $n_\mu \propto \pad_\mu \Phi$ satisfies
 $n_\mu n^\mu  \propto 2 dr_-/du_->0$,} and thus the surface is timelike. The same can be observed by finding that $k^2<0$, where $k=(1,r'_-,0,0,)$ is the tangent vector to the
  hypersurface $r=r_\sg$.
  During evaporation $r_+'<0$,  the parametrization $\Phi(x)=r-r_+(u_+)=0$ leads to $\pad_\mu \Phi\pad^\mu\Phi=-2 dr_+/du_+>0$.

 When the energy density  is given by $\Xi=\Upsilon^2>0$, the Schwarzschild sphere forms only ``beyond the end of time.''
Taking $s=-1$,  a formal transformation to $(u_-,r)$ coordinates gives
\begin{align}
&\pad_{u_-} C_-\approx-8\pi r_-^2e^{h_-}\Upsilon^2, \\
&\pad_r C_-\approx0, \qquad  \pad_r h_-\approx 0.
\end{align}
Their solutions are
\begin{align}
&C_-=r_-+\cO(r-r_-)^2, \qquad h_-=\cO(r-r_-)^2,
\end{align}
with the higher-order terms determined by the full energy-momentum tensor. The leading terms correspond to the pure outgoing Vaidya metric with $C_-'<0$.
{Likewise,  if $s=+1$, we obtain the
ingoing Vaidya metric with $C_+'>0$  {up} to leading order in $(r-r_+)$.
The four cases of the limiting form of $T_{\mu\nu}$ are summarized in Table 1.
\begin{table}[!htb] \centering
	\begin{tabular}{c|c|c|c} \hline \small  		\begin{tabular}{@{}c@{}}  $\mathrm{sgn}(T_{tt})$ \\ 
		\end{tabular}
		&
		\begin{tabular}{@{}c@{}} $\mathrm{sgn}(T_t^r)$  \\ 
		\end{tabular} 		
		&
		$s$
		&
		\begin{tabular}{@{}c@{}} {Vaidya metric} \\ 
		\end{tabular}
	\\ \hline \hline
	$-$ & $-$ & $+$ & \hspace*{-5mm} $(u_+,r)$ \quad $C^\prime(u_+) < 0$ \hspace*{-5mm} \\ \hline
	$-$ & $+$ & $-$ &\hspace*{-5mm}  $(u_-,r)$ \quad $C^\prime(u_-) > 0$ \hspace*{-5mm} \\ \hline
	$+$ & $-$ & $-$ & \hspace*{2.75mm} $(u_-,r)$ \quad $C^\prime(u_-) < 0$ \hspace*{2mm}
	\tabularnewline \hline
	$+$ & $+$ & $+$ & \hspace*{-5mm} $(u_+,r)$ \quad $C^\prime(u_+) > 0$ \hspace*{-5mm} \tabularnewline \hline
	\end{tabular}
	\caption{\small Signs in the limiting form of $T_{\mu\nu}$. The Einstein equations have real solutions only in the first two cases.
	} 	\label{Tab:Signs} \vspace{-2mm}
\end{table}

 In these coordinates the existence of an  {apparent}   {horizon radius} $r_\pm$ is compatible with
positive energy density. However, since  violation of the energy conditions is a coordinate-independent property,  there is no transformation between $(u_\pm,r)$
coordinates to $(t,r)$ coordinates when $\Xi>0$.
    Indeed, when such a transformation is constructed (Appendix~C), it results
in complex-valued functions $t(u_\pm,r)$.

If a trapped region appears at finite time $t_\mathrm{S}$ of Bob, it can be maintained only if the total matter
content outside it violates  {the} NEC and thus all energy conditions.
This must hold regardless of the presence or absence of the Hawking-like radiation that is observable by Bob.  As a result, the question of the existence of a BH
(that we take to be an ABH  with a trapped region inside, with or without singularities) becomes the question of the possibility of maintaining the required violation of
the energy conditions \cite{mmv:17,few:17} for  {a} sufficient  {amount of} time.

We now consider the implications of
quantum energy inequalities (QEI)  for our results.  These inequalities generalize the classical positive energy conditions and arise because classical energy conditions do not hold pointwise in a quantum field theory \cite{few:17}.  They
necessarily invoke   time averages, as there are no explicit
restrictions on spatial averages \cite{fhr:02}. We assume that the lifetime of the evaporating BH is predicted by the usual analysis \cite{fn:book,rbm:book}, i.e.,
by $\Delta t_\mathrm{E} \propto M_0^3$, where $M_0$ is the
 gravitational
 mass at the appearance of the Schwarzschild sphere. 

 The rigorous answer can be obtained by adapting the inequalities bounding violation  of the NEC \cite{few:17,fp:06}  to the relevant near-horizon metrics.
   There is a one-to-one relationship between  a static observer near an event horizon of a Schwarzschild BH and a uniformly accelerated observer in Minkowski spacetime.
 We exploit its approximate     applicability in our time-dependent scenario \cite{p:14}.

   Reference \cite{fp:06} established a QEI for an observer
moving with   uniform acceleration $w^\mu$ on the trajectory $\gamma$ that is parametrized by the proper
time $\tau$ in Minkowski spacetime (or a spacetime that can be mapped to Minkowski under a precise set of conditions). Then for any Hadamard state $\omega$ and a smooth
function $\mathfrak{f}(\tau)$ of compact support,
$\int_{-\infty}^\infty \mathfrak{f}^2(\tau)d\tau=1$, the negativity of the energy-momentum expectation value is bounded by
\be
\lim_{\tau_0\to\infty} \inf\frac{1}{\tau_0} \int_\gamma \mathfrak{f}(\tau/\tau_0)^2\6\hat T_{\mu\nu}\9_\omega u^\mu u^\nu d\tau \geqslant -\frac{11 w^4}{480\pi^2} \label{qei}
\ee

We  apply this bound to a stationary observer Eve in the vicinity of the Schwarzschild sphere of a slowly evaporating BH.  In $(u_+,r)$
coordinates the metric is given by Eq.~\eqref{metdec} and her four-velocity by $u^\mu_\mathrm{Eve}=(1/\sqrt{f},0,0,0)$. Thus Eve's energy density is
\be
\rho_\mathrm{Eve}= T_{\mu\nu} u^\mu_\mathrm{Eve} u^\nu_\mathrm{Eve} \approx-\frac{\Upsilon^2}{f}\approx\frac{C'_+}{8\pi C_+ x}.
\ee
It is useful to introduce the reduced variables $\chi \vcentcolon= x/C_+$ and $\beta \vcentcolon= -C'_+/C_+$. The near-horizon approximation is valid for $\chi\ll 1$, while  the Planck scale corresponds to $x=\chi C_+=1$.
 For Eve at rest close to the Schwarzschild sphere, the fourth power of the four-acceleration satisfies
\be
 w_\mathrm{Eve}^4= \frac{(\chi-\beta C_+(1+\chi)^2)^4}{16 C^4_+\chi^6(1+\chi)^6},
\ee
where for a macroscopic BH we expect $\beta\ll 1$. This quantity diverges as $\chi\to 0$, but it sharply drops to zero and stays close to it in a narrow interval at the sub-Planckian value
\be
\xi_0=\beta C_+, \ee
while $\rho_\mathrm{Eve}\approx \beta/(8\pi \xi_0)$.
We expect Eq.~\eqref{qei} to be applicable if it is possible to have $\tau_0$ long enough, i.e., $w_\mathrm{Eve}\tau_0 \gg 1$, which is still short enough on the evaporation scale, $ \Delta u_+\ll u_+^\mathrm{E}$. The proper time is obtained via  $d\tau=\sqrt{f}dv$.
 Then   (assuming that Eve is outside of the sub-Planckian region, i.e., $\chi\gg \beta C_+$) \be
\tau_0\approx\sqrt{\frac{x}{C_+}} \Delta u_+ =\sqrt{\chi}\Delta u_+\gg 2C_+\sqrt{\chi}, \label{shortau}
\ee
where  $C_+$ is approximately constant during $\tau_0$, and the double bound is
\be
u^\mathrm{E}_+\gg \Delta u_+\gg C_+,
\ee
which can be satisfied for a macroscopic BH.  The estimate does not change when $\chi\sim \beta C_+$. Approximating  the sampling function $\mathfrak{f}^2$  by  a box of width $\tau_0$,
 we  approximate the left-hand side of Eq.~\eqref{qei} as $\rho_\mathrm{E}$.

{Comparing it with $w_\mathrm{Eve}^4$ we find that the continuous domain where the QEI is satisfied belongs to the sub-Planckian regime, while}
 calculations that include backreaction of the Hawking radiation energy-momentum tensor on the geometry lead to  Eq.~\eqref{metdec}
as the leading correction valid for distances that are comparable with $r_\sg$ \cite{bmps:95}. Hence the above estimate can indicate either that the required negative energy density for
having a  Schwarzschild sphere at finite time $t_\mathrm{S}$ cannot be maintained, or that the Vaidya approximation of Eq.~\eqref{metdec} to the near-horizon geometry is valid in a much narrower
regime than previously thought. The latter may be another indication that the semiclassical approximation and its associated classical notions are modified already at the horizon or larger scales.

If we ignore this narrow sub-Planckian feature of $w_\mathrm{Eve}^4$, we see that the QEI is satisfied at larger scales, but the extent of the possible NEC-violating region
 depends on the details of evaporation. This can be seen as follows.  For a sufficiently slow evaporation ($\beta C^2\ll 1$, which we expect to be true for macroscopic BHs),  and
  for distances above the Planck scale $\chi\gtrsim 1/C_+$, we can approximate
 $w_\mathrm{Eve}^4$ by setting $\beta=0$. Then the QEI becomes
\be
\frac{11}{960C_+^4\pi \chi(1+\chi)^6} \geqslant\beta.
\ee
For $\beta C_+^4\lesssim 1$ it is satisfied at macroscopic scales, but with increasing (even still satisfying $\beta C_+^3\ll 1$) evaporation rates
\be
\chi < \frac{11}{960\pi \beta C_+^4 },
\ee
 again pushing the NEC-violating regime to the Planck scale.

 \section{Discussion}

 We demonstrated that having {a trapped region appear in finite time from the perspective of a distant observer} not only leads to {the} violation of
 energy conditions but, together with {the} requirement of regularity of its boundary, determines the limiting form of the total energy-momentum tensor and the metric near the Schwarzschild sphere.
 {This} physically motivated classification of the energy-momentum tensor near $r_\sg$ and the explicit expression for the metric allows to resolve the controversies that surround the thin shell models of collapse \cite{bmt-5}.

If the singularity forms at some finite $t$  (as  may happen even if some energy conditions are violated \cite{fg:11}),
 a version of the information loss problem may be posited \cite{visser, rbm:book}. The firewall paradox plays a prominent role among its numerous aspects.
Our analysis clarifies what the ``no drama at the horizon'' postulate that is used in its derivation actually means. So long as near the Schwarzschild sphere the curvature remains finite, the
resulting average energy density as perceived by the infalling Alice is negative, as can be seen from Eq.~\eqref{tneg}. Hence she does not see the vacuum, which is one of the elements in the chain of arguments at
least in some versions of the firewall paradox \cite{rbm:book}.
Moreover, if the  flux of energetic particles is desired to disentangle the early and late modes of emitted radiation, it has  {to not} only avoid being a universal disentangler (forbidden by unitarity) \cite{bht:17},
but also  be weak enough so as  {to not} destroy the horizon while being energetic enough for the erasure of information (at or above the Landauer bound) \cite{landauer}.  This is a highly nontrivial requirement.
Indeed, such a burst may  {still be} inefficient in destroying the correlations while causing the renormalized energy density $\6\hat{T}_{00}\9$  {to} diverge to positive infinity in the energy pulse \cite{cklz:18}.

Several problems present themselves as the logical extension of this work: (i) existence of trapped regions at finite $t$  with axial symmetry;
(ii) rigorous derivation of the QEI for the resulting geometries and study of their implications on BH existence;  this will also include evaluation of the limiting value $g_0(t)$; (iii) investigation of the Schwarzschild radius,
energy conditions and metric in alternative theories of gravity;
 (iv) finding possible observational differences with the black holes of GR, particularly via quasinormal modes.

\acknowledgments
We thank Amos Ori and Paddy Padmanabhan  for important critical comments and Mark Wardle for useful discussions on astrophysical black holes. The work of RBM was supported in part by the Natural Sciences and Engineering Research Council of Canada. SM is supported by the iMQRES scheme of Macquarie University.

\appendix

 \section{Metric near $\boldsymbol{r_\sg(t)}$ and different limiting forms of the energy-momentum tensor } \label{apA}

 \subsection{Regularity of $\mathbf T^\theta_\theta$}\label{apreg}

If we assume that  $T^\theta_\theta$ diverges at $r_\sg$, then the finiteness of the scalars $T$ and $\mathfrak{T}$ requires that it also has the asymptotic form $\Omega/f$. As a result  the regularity conditions become
\be
-\Xi+\Psi+2 \Omega=0, \qquad -2\Sigma^2 +\Xi^2+\Psi^2+2\Omega
^2=0,
\ee
where
\begin{align}
&\lim_{r\to r_\sg} \tau^r=\Psi(t), \\ &\lim_{r\to r_\sg} \tau_t=\Xi(t), \\ &\lim_{r\to r_\sg} \tau_t^r=\Sigma(t).
\end{align}
Taking $\Xi$ and $\Psi$ as independent variables we find
\be
\Omega=\half(\Xi-\Psi), \qquad \Sigma^2=\frac{1}{4}(3 \Xi^2+3\Psi^2-2\Xi\Psi).
\ee

The Einstein equation \eqref{gtt} does not change; its solution is
\be
C=r_\sg+W(t,x)=\frac{2}{3}r_\sg-4\Upsilon_\Xi\sqrt{\pi r_\sg^3(r-r_\sg)}+\frac{1}{3}r\ldots,
\ee
where $\Xi=-\Upsilon_\Xi^2$. However, Eq.~\eqref{grr} becomes
\be
\pad_x h=4\pi\frac{(\Sigma-\Upsilon_\Xi^2)}{(x-W)^2}.
\ee
Its solution is
\be
h=h_0(t)+\frac{\Sigma-\Upsilon_\Xi^2}{4\Upsilon_\Xi^2}\ln\big(r/r_\sg-1\big)+\ldots.
\ee

Hence, on the one hand,
\be
 \pad_t C =\frac{2 \Upsilon_\Xi\sqrt{\pi r_\sg^3}}{\sqrt{r-r_\sg}} \dot r_\sg+\ldots,
\ee
and it can be matched with
 \be
\frac{\pad_t C  }{r^2}=8\pi T_t^{\,r}\approx \pm 8\pi\Sigma e^{h}
\ee
only if $h\propto x^{-1/2}$, implying $\Psi=\Xi=-\Upsilon^2$. Hence   $\Omega=0$ and $T^\theta_\theta$ is finite at the Schwarzschild sphere.

\subsection{The Ricci scalar} \label{apR}

The explicit form of the Ricci tensor is given by
 \begin{align}
Rr^2=& -(4r-C) \pad_r h-2 r(r-C)(\pad_r h)^2\nonumber  \\
&+\pad_rC(2+3r\pad_r h)+r\pad_r^2C-2r(r-C)\pad_r^2h  \nonumber  \\
&+\frac{e^{-2h} r^3}{(r-C)^3} \big(2(\pad_t C )^2-(r-C)\pad_t C\pad_t h \nonumber \\
&+(r-C)\pad_t^2C\big).
\end{align}
Expansion in powers of $x=r-r_\sg$ results in the Laurent series
\begin{align}
R &=\varrho_3 x^{-3/2}+\varrho_2 x^{-1}+\varrho_1 x^{-1/2}+\cO(x) \nonumber \\
 &=\frac{p_3}{8\sqrt{\pi r_\sg}\xi_0^2\Upsilon}\frac{1}{x^{3/2}}+\frac{p_2}{16 \pi r_\sg^2\xi_0^2\Upsilon^2}\frac{1}{x} \nonumber \\
&\qquad +\frac{p_1}{144\pi^{3/2}\xi_0^3r_\sg^{7/2}\Upsilon^3}\frac{1}{\sqrt{x}}+\cO(x^0),
\label{rhoR}
\end{align}
in $\sqrt{x}$, with
\begin{align}
p_3 &=(\dot r_\sg^2-16\pi \xi_0^2r_\sg\Upsilon^2), \\
p_2& =\dot r_\sg^2-16\pi \xi_0^2r_\sg\Upsilon^2, \\
p_1&=-32\pi \xi_0^3 r_\sg\Upsilon^2-18\pi\dot\xi_0\dot r_\sg r_\sg^3\Upsilon^2 \nonumber \\
&+\xi_0[(2-9\pi r_\sg^2\Upsilon^2)\dot r_\sg^2-18\pi r_\sg^3\Upsilon(\dot r_\sg\dot \Upsilon-  {\Upsilon}\ddot r_\sg)].
\end{align}
where the denominators are the functions $\rho_i$ that were defined in Sec.~\ref{se2}.

\subsection{ The case  $\mathbf{\Xi=0}$} \label{apXi}

Here we discuss the case of the limits in Eq.~\eqref{limits} being zero.
Continuity of $T$ and $\mathfrak{T}$  does not impose any additional restrictions on their behavior; we summarize this situation as
 \begin{align}
 &\lim_{r\to r_\sg}\frac{e^{-2h }T_{tt}}{f ^k}=A(t), \\ &\lim_{r\to r_\sg}\frac{T^{rr}}{f ^l}=B(t),  \\  &\lim_{r\to r_\sg}\frac{e^{-h }T_{t}^{\,r}}{f ^m}=L(t),
 \end{align}
 for some $k,l,m\geq 1$, where \textit{a priori} there is no relationship between the functions $A$, $B$ and $L$.

Consider first the simplest case $k=1$.   For regular functions we have to take into account the next term in the expansion of the $\tau_t$ and other components in $r-r_\sg$.
 Hence close to $r_\sg$  Eq.~\eqref{gtt}  implies
 \be
  \pad_r C\approx8\pi \big(A  +A_1(r-r_\sg)\big){r^2_\sg},
  \ee
 where $A_1 \vcentcolon= 3 r_\sg^2 A$.
Its solution is
 \be
 C=r_\sg+8\pi Ar_\sg^2(r-r_\sg)+(4\pi A_1 r_\sg^2)(r-r_\sg)^2 +\ldots, \ee
where     the requirement $W<x=r-r_\sg$ for $x>0$ leads to the inequality
\be
8 \pi Ar_\sg^2\leq 1.
\ee

If the inequality is strict, i.e., $8 \pi Ar_\sg^2<1$, then the function $h$ either diverges at least logarithmically due to
\be
\pad_r h \approx  \frac{4\pi(A+B)r^2}{r-C}\approx \frac{4\pi(A+B)r_0^2}{(1-8\pi Ar_\sg^2)(r-r_\sg)},
\ee
or is regular if $A=-B$.  Matching the two expressions for $\pad_t C$  {at $r=r_\sg$} requires
\be
 \pad_t C =(1-8 \pi Ar_\sg^2)\dot r_\sg =r_\sg^2L\lim_{r\to r_\sg}  e^h f^m.
\ee
This identity can be satisfied only if $h$ diverges logarithmically  with a coefficient $\mu=-1$ while $m=1$.  Hence  (in  a gauge where $\xi_0=\sqrt{r_\sg}$) 
\be
\frac{4\pi(A+B)r_\sg^2}{(1-8\pi Ar_\sg^2)}=-1, \qquad \dot r_\sg=\frac{L r_\sg^2}{(1-8\pi Ar_\sg^2)}
\ee
We find that $R$ and $\mathfrak{R}$  are finite at $r_\sg$ if an additional condition
 \be
(1-8 \pi Ar_\sg^2)^2=\dot r_\sg^2
\ee
is satisfied, relating $A$ and $L$.

On the other hand, if  $8 \pi Ar_\sg^2=1$, then
\be
C=r-\beta^2 x^2,
\ee
 where $\beta^2=-(8\pi Ar_\sg+4\pi A_1 r_\sg^2)$. In this case $\pad_t C_0=0$. The    function $h$ either diverges at least as $1/x$ due to
 \be
\pad_r h \approx  \frac{4\pi(A+B)r^2}{r-C_0}\approx -\frac{4\pi(A+B)r_0^2}{\beta^2 x^2},
\ee
 or is regular if $A=-B$.  Since $\exp (1/x)$ diverges faster than any inverse power of $x$, the identity
 \be
0=\pad_t C  =r_\sg^2L\lim_{r\to r_\sg}  e^h f^m
\ee
can be satisfied only   if  $A=-B$ (thus getting a finite $h$) and $m\geq 1$.  Nevertheless, the Ricci scalar diverges, unless both $\pad_t C_0$   and $\pad_t^2 C_0$ are zero. Hence this case is excluded as it is not
compatible with the evolution of the Schwarzschild radius.
As a result, the simplest case of $\Xi=0$ is given by
$ k=l=m=1$, and
\begin{align}
B&=\frac{-1+4A \pi r_\sg^2}{4\pi r_\sg^2}=A-\frac{1}{4\pi r_\sg^2}, \\
 L&=\pm \frac{(1-8 \pi A r_\sg^2)^2}{r_\sg^2}, \qquad 1>8 \pi A r_\sg^2. \label{xi0k1}
\end{align}

In this case the $(tr)$ block of the energy-momentum tensor in the orthonormal frame is
\be
T_{\hat{a}\hat{b}}=     \begin{pmatrix}
A & L \\
L   & B
\end{pmatrix},
\ee
where $B(A)$ and $L(A)$ and the bound on $A$  are given in Eq.~\eqref{xi0k1}. Using these relations we find that both for positive and negative $A$
the NEC is violated ($T_{\hat{a}\hat{b}} k^{\hat{a}}k^{\hat{b}}<0$)
with the incoming radial null geodesic $k^{\hat{a}}=(1,-1,0,0)$.

 \section{$\boldsymbol{ (u_\pm,r)}$ coordinates and the Einstein equations: $\boldsymbol{\Xi<0}$} \label{apB}

 In this section we present the case of negative energy density when the partial differential equation {(PDE)} for the integrating factor $h_\pm$
 has a real-valued solution,
 meaning   that we are simply discussing the same physical situation in different coordinate systems. We denote the energy-momentum tensor in
 the two coordinate systems $(u_\pm,r)$ as
  $\Theta_{\mu\nu}^{\pm}$, respectively. To reduce clutter we restore $u=u_-$ and $v=u_+$.

 \subsection{$\boldsymbol{(u,r)}$ coordinates}

 First we analyze the retarded null coordinate.
 Note that {the Schwarzschild radius satisfies $r_-(u)=r_\sg\big(t(u,r_-(u))\big)$, so the numerical values of limits $\Upsilon$ are the same. Recall
\be
dt=e^{-h}\big(e^{h_-}du+dr/f).
\ee
 Hence the relevant components of the energy-momentum tensor for the two signs of the flux $\tau_t^r=-\Upsilon^2$ ($\tau_t^r=+\Upsilon^2$)   are
\begin{align}
\Theta_{uu}^{-}=&\left(\frac{\pad t}{\pad u}\right)^2T_{tt}\to -e^{2h_-}\Upsilon^2 \quad \big(-e^{2h_-}\Upsilon^2\big), \\
\Theta_{ur}^{-}=&\frac{\pad t}{\pad u}\frac{\pad t}{\pad r}T_{tt}+\frac{\pad t}{\pad u}T_{tr}\to-\frac{2}{f}e^{h_-}\Upsilon^2 \quad  \left(0\right),\\
\Theta_{rr}^{-}=&T_{rr}+\left(\frac{\pad t}{\pad r}\right)^2T_{tt}+2\frac{\pad t}{\pad r}T_{tr}\to-\frac{4}{f^2}\Upsilon^2 \quad \left(0\right).
\end{align}

In both cases the  left-hand sides of the Einstein equations are
\begin{align}
G_{uu}&=e^{2h_-}\frac{\pad_rC_-}{r^2}\left(1-\frac{C_-}{r}\right)-e^{h_-}\frac{\pad_uC_-}{r^2},\\
G_{ur}&=e^{h_-}\frac{\pad_r C_-}{r^2}, \\
G_{rr}&=\frac{2\pad_r h_-(u,r)}{r}. \label{hreq}
\end{align}


Close to the Schwarzschild sphere we get (for $\tau_t^r=-\Upsilon^2$, contraction of the apparent horizon)
\begin{align}
\pad_u C_-=&-8\pi r_-^2e^{h_-}\Upsilon^2, \\
\pad_r C_-=&-16\pi r_-^2\Upsilon^2/f, \\
\pad_r h_-=&-16\pi r_- \Upsilon^2/f^2.
\end{align}
The second and the third equation give
\be
C_-(u,r)=\tfrac{2}{3}r_-(u)-\sqrt{2}a(r-r_-)^{1/2}+\tfrac{1}{3}r\ldots,
\ee
and
\be
h_-=-\half\ln(r/r_--1) +\frac{2\sqrt{2}}{3}\frac{\sqrt{x}}{a} ,
\ee
where \be
a \vcentcolon= 4\sqrt{\pi}\,\Upsilon r_-^{3/2}. \ee

Similar to the analysis in $(t,r)$ coordinates  we have the identity between the two ways to calculate $\pad C_-(u,r_-)$.
On the one hand the Einstein equation implies
\be
\lim_{r\to r_-}\pad_u C_-(u,r)=-\frac{a^2}{2\sqrt{r_-(r-r_-)}},
\ee
and on the other hand
\be
\lim_{r\to r_-}\pad_u C_-(u,r)=\frac{a dr_-/du}{\sqrt{2(r-r_-)}}.
\ee
Hence
\be
\frac{dr_-}{du}=-\frac{a}{ \sqrt{2r_-}}.
\ee

\subsection{$\boldsymbol{(v,r)}$ coordinates}
The relevant components of the energy-momentum tensor for the two signs of the flux $\tau_t^r=-\Upsilon^2$ ($\tau_t^r=+\Upsilon^2$)   are \begin{align}
\Theta_{vv}^{+}=&\left(\frac{\pad t}{\pad v}\right)^2T_{tt}\to -e^{2h_+}\Upsilon^2 \left(-e^{2h_+}\Upsilon^2 \right), \\
\Theta_{vr}^{+}=&\frac{\pad t}{\pad v}\frac{\pad t}{\pad r}T_{tt}+\frac{\pad t}{\pad v}T_{tr}\to  0 \quad \left(\frac{2}{f}e^{h_+}\Upsilon^2 \right),\\
\Theta_{rr}^{+}=&T_{rr}+\left(\frac{\pad t}{\pad r}\right)^2T_{tt}+2\frac{\pad t}{\pad r}T_{tr}\to 0 \quad \left(\frac{4}{f^2}\Upsilon^2 \right).
\end{align}

The Einstein equations are
\begin{align}
G_{vv}&=e^{2h_2}\frac{\pad_rC_+}{r^2}\left(1-\frac{C_+}{r}\right)+e^{h_+}\frac{\pad_vC_+}{r^2},\\
G_{vr}&=-e^{h_+}\frac{\pad_r C_+}{r^2},\\
G_{rr}&=\frac{2\pad_r h_+(v,r)}{r}. \label{hreq}
\end{align}

 \section{$\boldsymbol{ (u_\pm,r)}$ coordinates and the Einstein equations: $\boldsymbol{\Xi>0}$}

The case $\Xi=\Upsilon^2$ with negative $\tau_t^r$ is most conveniently described using retarded null coordinates
\begin{align}
	\pad_u C_-&=-8\pi r_-^2\Upsilon^2  , \\
	\pad_r C_-&=0  , \\
	\pad_r h_-&=0  ,
\end{align}
which at leading order correspond to the standard outgoing Vaidya metric
\be
	ds^2=-\big(1-C_-\big)du_-^2 - 2 du_- dr + r^2d\Omega,
\ee
with ${C'_-}< 0$. Similarly, the case of positive energy density and positive $\tau_t^r$ is naturally described using advanced null coordinates,
resulting in the standard ingoing Vaidya metric
\be
	ds^2=-\big(1-C_+\big)du_+^2 + 2 du_+ dr + r^2d\Omega,
\ee
with ${C'_+} > 0$.\\
Equation \eqref{intf}, which relates the differential of $t$ to the differentials of $u_\mp$ and $r$, becomes
\begin{align}
	&dt = e^{-\psi_\mp(u_\mp,r)}(du_\mp \pm f^{-1} dr), \\ &f = 1 - \frac{C_\mp(t(u_\mp,r),r)}{r} = 1 - \frac{C_\mp(u_\mp)}{r} \; ,
\end{align}
where the partial differential equation (PDE) for the integrating factor $h=\psi_\mp(u_\mp,r)$  is
\be
	\partial_r \psi_\mp \mp f^{-1} \partial_{u_\mp} \psi_\mp =\pm f^{-2} \partial_{u_\mp} f . \label{eq:PDE}
\ee

In both cases we obtain complex-valued solutions for the  transformation from  $(u_\pm,r)$ to $(t,r)$ coordinates.
This is a manifestation of the fact that if the NEC is satisfied, there is no finite-time solution that has the Schwarzschild radius $r_\sg(t)$. First we recall some general properties of a first order {PDE}.

Using the standard procedure for solving a first order linear PDE \cite{pde1,pde2}, we set the characteristic system of ordinary differential equations
\begin{align}%
&\frac{dr}{a} = \frac{du_\mp}{b} = \frac{d\psi_\mp}{c}, \nonumber \\
& a = 1 \; , \; b = \mp f^{-1} \; , \; c = \pm f^{-2} \partial_{u_\mp} f =\mp f^{-2}C_\mp'/r.
\end{align}
When it is unambiguous we use $\psi=\psi_\mp(u_\mp,r)$, $z=u_\pm$, and $C=C_\mp$ to simplify the notation in what follows. Two independent equations can be selected by choosing, e.g., $z$ as the independent variable. Then the characteristic curves are given by $r=r(z,K_1)$ and $\psi=\psi(z,K_1,K_2)$, where $K_1$ and $K_2$ are constants.
The general solution is then implicitly given by
\be
F\big(K_2(z,K_1(z,r),\psi)\big)=0, \label{psieq}
\ee
where $F(y)$ is an  arbitrary function of a single argument and $K_1$, $K_2$ are expressed as functions of $z$, $r$ and $\psi$.
Since we are interested in $\psi$ solely for its capacity as an integrating factor, we take the simplest form of $\psi$ that is compatible with Eq.~\eqref{psieq}.

We select two independent equations and use the coordinate $x=r-C(z)$ to obtain the system
\begin{align}
	\frac{dr}{dz} & = \frac{dC}{dz} + \frac{dx}{dz} =\mp\left(1-\frac{C}{C+x} \right) =\mp \frac{x}{C+x}, \label{eq:ODE_dr/du_2} \\
	\frac{d\psi}{dz} & = \frac{1}{fr} \frac{dC}{dz} = \frac{C'}{x}, \label{eq:ODE_dpsi/du_2}
\end{align}
where the upper sign corresponds to $z=u_-$. To illustrate the key points it is enough to consider early stages of the evolution after formation of the Schwarzschild sphere.  In both cases
 the mass is a linear function of $z$,
\be
C_\pm(u_\pm)=C_0(1\pm\beta u_\pm) ,
\ee
where $\beta>0$ is a constant and $C_0$ is the BH mass on the appearance of the Schwarzschild sphere. The evaporating case  is described in $(u_-,r)$ coordinates, and the  case of growing BH mass  is described in $(u_+,r)$ coordinates.  We also consider only regions that
 are close to the Schwarzschild radius, $x\ll C$. The terms $\beta z C_0$ and $x$ are of the same order of magnitude, and we can keep only the first order expression when expanding the denominator of Eq.~\eqref{eq:ODE_dr/du_2}.

Consider first the outgoing Vaidya metric. Solving Eq.~\eqref{eq:ODE_dr/du_2}   we obtain (in the leading order)
\be
	K_1 = u_-+C_0 \ln\left(\frac{x}{C_0}-\beta C_0\right).
\ee
Substituting $x(u_-)$ into Eq.~\eqref{eq:ODE_dpsi/du_2} and integrating, we find
\be
K_2=\psi_-+\ln\big(e^{K_1/C_0}+\beta C_0 e^{u_-/C_0}\big).
\ee

\vspace*{2mm}
For the ingoing Vaidya metric Eq.~\eqref{eq:ODE_dr/du_2} gives
\be
	K_1 = u_+-C_0 \ln\left(\frac{x}{C_0}-\beta C_0\right)
\ee
and
\be
K_2=\psi_+-{u_+}/{C_0}+\ln\big(\beta C_0+e^{(u_+-K_1)/C_0}\big).
\ee
We see that $\psi_\pm(u_\pm)$ must be complex valued as expected for complex $C$ and $h$ in the case of $\Xi=\Upsilon^2$. On the other hand,
the case of $\Xi=-\Upsilon^2$ and  positive incoming flux results in a real-valued transformation that connects two coordinate representations of the same physical situation.

\end{document}